\documentclass[twocolumn,prl,showpacs,amsmath,amstex,amssymb,mathfonts,superscriptaddress]{revtex4-1}

\usepackage{amsthm,color,amsfonts,graphicx,verbatim}
\usepackage{tikz}
\usepackage{amsmath}
\usepackage{amssymb}
\usepackage{amsthm}
\usepackage{amsfonts}
\usepackage{listings}
\usepackage{enumerate}
\usepackage{latexsym}
\usepackage{psfrag}
\usepackage{bm}
\usepackage{graphicx}
\usepackage{dsfont}

\newcommand{\be}{\begin{equation}}
\newcommand{\ee}{\end{equation}}
\newcommand{\bea}{\begin{eqnarray}}
\newcommand{\eea}{\end{eqnarray}}

\newcommand{\la}{\langle}
\newcommand{\ra}{\rangle}

\renewcommand{\phi}{\varphi}
\renewcommand{\epsilon}{\varepsilon}

\begin{document}

%FW commands
\newcommand{\C}{\mathbb{C}}
\newcommand{\R}{\mathbb{R}}
\newcommand{\Q}{\mathbb{Q}}
\newcommand{\Z}{\mathbb{Z}}
\newcommand{\N}{\mathbb{N}}
\newcommand{\T}{\mathbb{T}}

\newcommand{\dd}{\mathrm{d}}
\newcommand{\id}{\mathrm{i}}
\newcommand{\ed}{\mathrm{e}}

\newcommand{\caA}{{\mathcal A}}
\newcommand{\caB}{{\mathcal B}}
\newcommand{\caC}{{\mathcal C}}
\newcommand{\caD}{{\mathcal D}}
\newcommand{\caE}{{\mathcal E}}
\newcommand{\caF}{{\mathcal F}}
\newcommand{\caG}{{\mathcal G}}
\newcommand{\caH}{{\mathcal H}}
\newcommand{\caI}{{\mathcal I}}
\newcommand{\caJ}{{\mathcal J}}
\newcommand{\caK}{{\mathcal K}}
\newcommand{\caL}{{\mathcal L}}
\newcommand{\caM}{{\mathcal M}}
\newcommand{\caN}{{\mathcal N}}
\newcommand{\caO}{{\mathcal O}}
\newcommand{\caP}{{\mathcal P}}
\newcommand{\caQ}{{\mathcal Q}}
\newcommand{\caR}{{\mathcal R}}
\newcommand{\caS}{{\mathcal S}}
\newcommand{\caT}{{\mathcal T}}
\newcommand{\caU}{{\mathcal U}}
\newcommand{\caV}{{\mathcal V}}
\newcommand{\caW}{{\mathcal W}}
\newcommand{\caX}{{\mathcal X}}
\newcommand{\caY}{{\mathcal Y}}
\newcommand{\caZ}{{\mathcal Z}}

\newcommand{\Const}{\mathrm{C}}
\newcommand{\const}{c}
\newcommand{\8}{\infty}
\newcommand{\str}{ |}
\newcommand{\e}{ \mathrm{e}}
\newcommand{\iu}{ \mathrm{i}}
\newcommand{\norm}{ \|}
\newcommand{\beq}{ \begin{equation} }
\newcommand{\eeq}{ \end{equation} }
\newcommand{\Res}{\mathsf{Res}}
\newcommand{\Proba}{\mathsf{P}}
\newcommand{\ad}{\mathrm{ad}}

\definecolor{Green}{rgb}{0.2,0.7,0.1}
\definecolor{Red}{rgb}{0.9,0.0,0.1}

% Comments
\newcommand{\wdr}{\textcolor{blue}}
\newcommand{\fh}{\textcolor{Green}}
\newcommand{\fhB}{\textcolor{magenta}}
\newcommand{\da}{\textcolor{purple}}

\title{A theory of many-body localization in periodically driven systems}

\author{Dmitry A. Abanin}
\affiliation{{Department of Theoretical Physics, University of Geneva, on leave from}}
\affiliation{Perimeter Institute for Theoretical Physics, Waterloo, Canada}
\author{Wojciech De Roeck}
\affiliation{Instituut voor Theoretische Fysica, KU Leuven, Belgium}
\author{Fran\c{c}ois Huveneers}
\affiliation{CEREMADE, Universit\'e Paris-Dauphine, France}

\date{\today}

\begin{abstract}
We present a theory of periodically driven, many-body localized (MBL) systems. 
We argue that MBL persists under periodic driving at high enough driving frequency:
The Floquet operator (evolution operator over one driving period) can be represented as an exponential of an effective time-independent Hamiltonian, which is a sum of quasi-local terms and is itself fully MBL.
We derive this result by constructing a sequence of canonical transformations to remove the time-dependence from the original Hamiltonian. 
When the driving evolves smoothly in time, the theory can be sharpened by estimating the probability of adiabatic Landau-Zener transitions at many-body level crossings.
In all cases, we argue that there is delocalization at sufficiently low frequency. 
We propose a phase diagram of driven MBL systems. 
%The Floquet eigenstates in this case have area-law entanglement entropy, and there exists an extensive set of local integrals of motion. 
%We argue that at sufficiently low frequency, there is always delocalization, owing to a large number of many-body level crossings and non-diabatic Landau-Zener transition between them. 
\end{abstract}

\maketitle

% SECTION
{\bf Introduction.} 
Recently, there has been much interest in quantum many-body localized (MBL) systems and their properties~\cite{andersonabsence, juergandspencer83, Basko06,Mirlin05,Oganesyan07,Prosen08,Pal10,Moore12,Vosk13,Serbyn13-1, Huse13-1, Serbyn13-2,Huse13,bauer, Imbrie,Pekker14,Kjall14,Ros14,Chandran14,Serbyn14,Vasseur14,Khemani14}. 
MBL phase is characterized by an extensive set of emergent local integrals of motion (LIOMs)~\cite{Serbyn13-2,Huse13}, which lead to quantum ergodicity breaking, and in particular, absence of thermalization. 
Therefore, MBL systems cannot be described by conventional statistical mechanics. 
Existing works explored experimental manifestations of MBL systems, and predicted universal dynamical properties following a sudden quantum quench, including logarithmic growth of entanglement entropy~\cite{Prosen08,Moore12,Vosk13,Serbyn13-1,Serbyn13-2,Huse13}, as well as characteristic decay~\cite{Serbyn14} and revivals~\cite{Vasseur14} of local observables. 

In this paper, we study the behaviour of MBL systems under periodic driving. 
Previous works on driven many-body systems focused mostly on the translationally invariant case~\cite{Prosen98,Silva12,Alessio13,Alessio14,Mueller14,Lazarides14-1}. 
In particular, D'Alessio and Polkovnikov~\cite{Alessio13} conjectured that, if the dynamics is generated by switching between an ergodic and an integrable (but translationally-invariant) Hamiltonian, a transition will be observed in function of the driving frequency: 
At low frequency, the system shows heating to an infinite temperature, while at high frequency the dynamics is described by an effective Hamiltonian written as a sum of local terms, %so that there is an integral of motion in the system, 
leading to localization in the energy space. 
{Though very long time scales can indeed be needed for energy to get dissipated \cite{Aba DeR Huv}, 
it was argued that driven ergodic systems typically delocalize and heat up to an infinite temperature at any driving frequency  \cite{Alessio14,Lazarides14-1,Ponte14-1}.}
%For translationally invariant systems, this conjecture remains unproven and, moreover, it was argued that driven ergodic systems typically delocalize and heat up to an infinite temperature  \cite{Alessio14,Lazarides14-1,Ponte14-1}. 

There are three main motivations to our work. 
First, studying the response of many-body systems to periodically varying fields is a conventional experimental probe in systems of cold atoms in optical lattices~\cite{Bloch08,Polkovnikov11}, 
which are promising candidates for realizing the MBL phase~\cite{Kondov13,Inguscio13}.
Second, theoretically little is known about general properties of quantum many-body systems under time-varying fields (beyond linear-response). 
{Finally, we investigate the conjecture in~\cite{Alessio13}, in the context of MBL systems, where counter-arguments based on ergodicity fail in an obvious way.}
%\fhB{I agree formulation was inaccurate, but at least for me, this counts as a motivation.}
%Finally, we show that, for the particular case of MBL systems, the conjecture of \cite{Alessio13} is correct: at high frequency, the effective Hamiltonian is a sum of local terms, and is actually itself MBL.

We consider a time-dependent periodic Hamiltonian $H(t)= H(t+T)$ and we split it in its mean and oscillating parts:
$H(t) = H^{(0)} + V(t)$ with $H^{(0)} = \frac{1}{T} \int^T_0 \dd t  \,  H(t)$.
%\begin{equation}\label{time dependent Hamiltonian}
%H(t) =  H^{(0)} + V(t), \qquad H^{(0)} = \frac{1}{T} \int^T_0 \dd t  \,  H(t).
%\end{equation}
We analyze the one-cycle evolution operator $U(T)=:\ed^{-i H_* T}$, where $H_*$ is an effective (Floquet) Hamiltonian, which a priori can be nonlocal, and where the evolution operator $U(\cdot)$ solves
\begin{equation}
\iu\frac{\dd}{\dd t} U(t)=H(t)U(t), \quad U(0)=1.
\end{equation}
We study the case where the time-averaged Hamiltonian $H^{(0)}$ is fully MBL (i.e.\@ has all its eigenstates localized), 
and determine conditions on $V(t)$ so that $H_*$ is still local and fully MBL which implies, in particular, energy localization \cite{remarkenergyloc}.
Our work consists of two parts: 

(a) 
We show that $H_*$ is MBL via successive canonical transformations. 
Our method is directly inspired {by} the scheme devised by Imbrie~\cite{Imbrie} (see also \cite{Bambusi}) to establish the existence of a localized phase for time-independent Hamiltonians. 
As this scheme allows to go beyond asymptotic expansions, we claim that it furnishes a more robust foundation to MBL in driven systems than the use of Magnus expansion~\cite{Alessio13}. 
We emphasize both that $H^{(0)}$ can be fully MBL even when the instantaneous Hamiltonian $H(t)$ is ergodic for most or even all $t \in [0,T]$, 
and that  $V(t)$ is not required to vary continuously with time (square signals as in~\cite{Ponte14-2} are allowed). 

(b) 
We assume that $V(t)$ involves only a few harmonics, i.e.\ varies smoothly with time. 
In that case, we don't expect the scheme mentioned in (a) to lead to optimal conditions on the lowest possible frequency to ensure MBL. 
Instead, we base our analysis on an analogy with the multi-level Landau-Zener problem. 
In particular, we argue that at sufficiently low frequency, the Floquet operator strongly mixes states with a very different spatial structure, thus inducing delocalization. 

\begin{figure}[htb]
\begin{center}

\begin{tikzpicture}[scale=1]

\draw[->,>=latex] (-0.3,0) -- (6,0);
\draw[->,>=latex] (0,-0.3) -- (0,4);

%\draw[dotted] (1,0) -- (1,1) -- (0,1);

\draw (1,0) -- (1,-0.1);
\draw (0,1) -- (-0.1,1);

\draw (1,0) node[below]{1};
\draw (0,1) node[left]{1};

\draw[ultra thick,Red] plot [domain=0.265:5.8,smooth] (\x, {1/\x}); 
%\draw[ultra thick,Red] plot [domain=0.13:1,smooth] (\x, {1/\x^(2/3)}); 

\draw  (0,3.5) node[left] {{$W/g$}};
\draw  (5.5,0) node[below] {{$\nu/g$}};

\draw  (0.9,0.15) node[above] {Thermal};
\draw  (3,2) node[above] {Localized};

\end{tikzpicture}

\caption{ \label{Fig3} 
Qualitative phase diagram of the driven MBL system. 
For $\nu/g > 1$ (and $W/g <1$), the transition is determined by \eqref{conditions on T and g}: $W/g \sim (\nu/g)^{-1}$.
For $\nu/g \le 1$ (and $W/g \ge 1$), and for a smooth driving, the transition is predicted by \eqref{eq:nu_critical}: $W/g \sim (\nu/g)^{-a}$ with $a\sim \xi$.
%Since $a \sim \xi$, it holds that $a\sim 1$ as $W/g \sim 1$ and $a \to 0$ as $W/g \to \infty$. \fh{Not correct because $\xi$ has some value from $H_0$.}
%\fh{In fact, it stays valid for $h< g< W$ but for $g<h$ it stops being true.} \fh{Perhaps no bump on the figure.}
}
\end{center}
\end{figure}
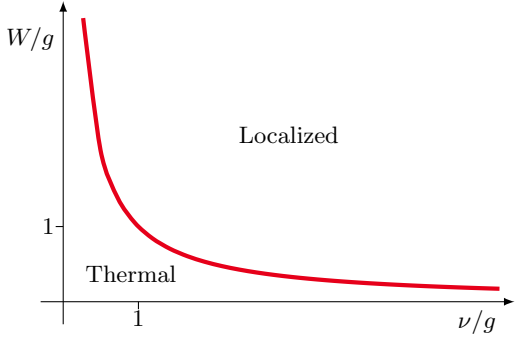

Based on (a) and (b), we propose a qualitative phase diagram of driven MBL systems (see Fig.~\ref{Fig3}). 

% SECTION
{\bf Model.} 
For concreteness, we assume that our system is a one-dimensional spin-$1/2$ chain of size $L$. We make the following assumptions:
(a) The Hamiltonian $H^{(0)}$ is fully MBL, and therefore it has a complete set of LIOMs. 
Choosing the LIOMs to be the local spins $\sigma_j^{z}$, %(third Pauli matrix), 
$H^{(0)}$ takes the form (see \cite{Serbyn13-2,Huse13,Imbrie})
\begin{multline}\label{eq: mbl ham}
H^{(0)} = \sum_{i} \epsilon_j \sigma^z_j +\sum_{i<j}  \epsilon_{i,j} \sigma_i^z \sigma_j^z + \dots  \\
 + \sum_{i_1<\dots <i_n} \epsilon_{i_1,\dots,i_n} \sigma_{i_1}^z \dots \sigma_{i_n}^z + \dots
\end{multline}
where  $|\epsilon_{i_1,\dots,i_n} | \sim \ed^{-( i_n - i_1)/\xi}$, except at rare resonant spots. 
Through this work we assume that $H^{(0)}$ is strongly localized: $\xi \ll 1$. 
% and where $\xi \lesssim 1$ is the localization length of $H^{(0)}$.
%(strictly speaking, due to rare resonant spots, the localization length $\xi$ should be allowed to vary form place to place).
(b) The energies $\epsilon_{i_1,\dots i_n}$ are functions of the local disorder, hence random; we set
\begin{equation}
\langle \epsilon_i \rangle = E_0, \quad \langle (\epsilon_i - E_0)^2 \rangle = W^2,
\end{equation} 
and we assume $E_0 \sim W$. 
(c) The driving is of the form 
\begin{equation}\label{eq: driving}
V(t) = \sum_{i} V_i(t), \; \frac{1}{T} \int_0^T \dd t \, V_i (t) = 0, \;  \| V_i(t) \| \sim g,
\end{equation} 
{where $g$ is the driving strength, and} where $V_i(t)=V_i(t+T)$ are local around site $i$, implying $\| [V_i(t),\sigma_j^{w}] \| \lesssim \ed^{-|i-j|/\xi}$ for $w=x,y,z$. 
%E.g. $V_i (t) = \cos (2\pi \nu t) \hat V_i$ for some local $\hat V_i$, with $\nu = 1/T$. 

%%% SECTION
\textbf{Localization at high frequency.}
%We assume $\xi \ll 1$. 
To get explicit expressions, we take $H^{(0)}$ to be given by the first sum in \eqref{eq: mbl ham} only \cite{neglect-original-res}, while for $V(t)$ we take the simple toy model 
\begin{equation}
V(t) =  \sum_{i} \sum_{w_1,w_2=x,y,z} J_{i,i+1}^{w_1,w_2}(t) \sigma_i^{w_1} \sigma_{i+1}^{w_2}
\end{equation}
with $J_{i,i+1}^{w_1,w_2}(t) = J_{i,i+1}^{w_2,w_1}(t)$ and
\begin{equation}\label{bound on V}
\sup_{t,i,w_1,w_2} |J_{i,i+1}^{w_1,w_2} (t)| \le g.
\end{equation}
It is one of our main observations that, for $\nu = 1/T$ large enough compared to $g$, the uniform bound \eqref{bound on V} is all we need: 
{as we argue now,} MBL in periodically driven systems can be understood in very much the same way as MBL in isolated systems.
%\fhB{I am a bit embarrassed by your remark since it is the aim that people realize that, \emph{afterwards}, it is indeed very much the same as static MBL. I tried a minimal modification.}
%(to substantiate this claim, we show in the Supplementary Material how to adapt the scheme by Imbrie \cite{Imbrie} to periodic in time Hamiltonians). 

{Let $H_*$ be such that $U(T) = \ed^{-i H_* T}$. }
There exists a periodic unitary $P(t)=P(t+T)$ such that $U(t) = P(t) \ed^{-\id H_* t}$; indeed we just define $P(t)$ by 
%Given $H_*$, there exists a periodic unitary $P(t)=P(t+T)$ such that $U(t) = P(t) \ed^{-\id H_* t}$
%(all effective Hamiltonians $H_*$ are simultaneously diagonalizable and their eigenvalues are related to one another by multiples of $2\pi \nu$; the choice of $H_*$ will not matter below). 
%Indeed we just define $P(t)$ by 
\begin{equation}\label{definition of P(t)}
P(t) = U(t)\ed^{\id H_* t}.
\end{equation}
%\fhB{The presentation can still look too convoluted, but I think it is really part of the logic to deal with resonances later on: we know that $H_*$ exists and that $P(t)$ can be inferred.}
This can be equivalently stated as 
%Since $P(0)=P(T)$ and since $P(t)$ obeys the evolution equation
%\begin{equation}\label{evolution equation P}
%\id  \frac{\dd P(t)}{\dd t} = H(t) P(t) - P(t)H_*,
%\end{equation}
%we deduce that $P(t)=P(t+T)$ for all $t$.
%We next observe that \eqref{evolution equation P} can be rewritten as 
\begin{equation}\label{canonical transformation}
P^\dagger (t) \Big( H(t) - \id \frac{\dd }{\dd t} \Big) P(t) = H_*. 
\end{equation}
We now change the point of view and we take \eqref{canonical transformation} as our starting point: 
we don't assume {to know} $H_*$ and we look for a periodic unitary $P(t)=P(t+T)$ satisfying $P(0)=1$ {such} that the right hand side of \eqref{canonical transformation} is time-independent. 
%It is then in turn straightforward to check that $H_*$ is an effective Hamiltonian in the sense of \eqref{effective Hamiltonian}.

Formally, \eqref{canonical transformation} is solved by successive approximations: $P(t) = \lim_{n\to\infty} P_1(t) \dots P_n(t)$.
We write $H_1(t)=H(t)$ and, for $n\ge 1$, we will determine $P_n(t)$ so as to make $H_{n+1} (t)$ ``as close as possible" to a time-independent Hamiltonian, where $H_{n+1} (t)$ is defined by 
\begin{equation}\label{H n+1 from H n}
P_n^\dagger (t)  \Big( H_n(t) - \id \frac{\dd }{\dd t}  \Big) P_n(t) = H_{n+1}(t) \quad (n\ge 1).
\end{equation}
%$$ H_n (t) =  \big(P_1(t)\dots P_{n-1}(t)\big)^\dagger \Big( H(t) - \id \frac{\dd }{\dd t}  \Big) \big(P_1(t)\dots P_{n-1}(t)\big),$$
If the procedure is successful, $H_n(t)$ becomes truly time-independent in the limit $n\to \infty$: $H_* = \lim_{n\to \infty} H_n(t)$.

%Before proceeding concretely, let us briefly comment on the fact that \eqref{definition of P(t)} and \eqref{canonical transformation} are two complementary point of views that prove both useful. 
%On the one hand, \eqref{canonical transformation} is a convenient expression to set-up the iterative scheme below and to deal with the perturbative part of the problem, i.e.\@ ignoring resonances. 
%On the other hand, it is not so clear at once how to solve \eqref{canonical transformation} when a non-perturbative analysis is required (resonant islands). 
%The whole point is that, at any step of the scheme, the analysis of resonances always boils down to a finite dimensional problem (with a dimension staying fixed while taking the thermodynamic limit). 
%We then resort to the abstract definition \eqref{definition of P(t)}  of $P(t)$.
%This strategy is the exact counter-part of the philosophy developed in \cite{Imbrie} for time-independent Hamiltonians:  
%in those places where perturbation theory failed, it was still possible to invoke that any finite-dimensional hermitian matrix is diagonalizable.  
%Because of this, we do claim that the present theory provides a more robust foundation to energy localization than studies based on Magnus expansion \cite{Alessio13}, 
%which convergence cannot be established, and probably fails, except in very exceptional cases (free systems). 

As an exemplary case, let us implement the first step of the scheme, i.e.\@ determine $P_1(t)$ and $H_2(t)$. 
As in \cite{Imbrie}, we decompose $V(t)$ as $ V(t) = V^{per}(t) + V^{res}(t)$, where $V^{res} (t)$ includes the resonant transitions, that cannot be treated in perturbation:
$$ V^{res}(t) = \sum_{(i,i+1)\in\Res} \; \sum_{w_1,w_2=x,y,z} J_{i,i+1}^{(w_1,w_2)}(t)\sigma_i^{(w_1)} \sigma_{i+1}^{(w_2)} .$$
where $\Res$ is a time-independent set containing bonds where resonances (possibly) occur, defined as follows.
Let us denote the eigenstates of $H^{(0)}$ by $|\eta\rangle$, corresponding thus to all possible configurations of up and down spins. 
We say that $(i,i+1)\in \Res$ if there exist two states $|\eta\rangle$ and $|\eta'\rangle$ as well as $w_1,w_2$ such that $\langle \eta' |Ê\sigma_i^{(w_1)}\sigma_{i+1}^{(w_2)}| \eta \rangle \ne 0$ and such that 
\begin{equation}\label{resonance condition}
\inf_{k\in \Z_0}| \Delta H^{(0)}_{\eta,\eta'} - 2 \pi k \nu | \le \delta \nu
\end{equation}
where $ \Delta H^{(0)}_{\eta,\eta'} = \langle \eta' | H^{(0)} | \eta' \rangle -  \langle \eta | H^{(0)} | \eta \rangle$ and where $\delta \ll 1$ is some (partially arbitrary) threshold.
The probability that $(i,i+1)\in \Res$ is bounded as
\begin{equation}\label{probability resonance}
\Proba ((i,i+1) \in \Res) \le \Const \delta, 
\end{equation}
independently of $\nu$ and $W$ (see Supplementary Material~I).

We now proceed to the perturbative analysis: We decompose $P_1(t)$ as $P_1(t) = P_1^{res}(t) P^{per}_1 (t)$ and we determine $P^{per}_1(t)$. 
We define it as $P^{per}_1(t) = \ed^{A_1(0)} \ed^{-A_1(t)}$ where $A_1(t)$ is a dimensionless anti-hermitian matrix satisfying
\begin{equation}\label{perturbative equation} 
[H^{(0)},A_1(t)] - \id \frac{\dd A_1(t)}{\dd t} = V^{per} (t)
\end{equation}
with $A_1(t) =A_1(t+T)$.
Since $H_0$ is MBL and since \eqref{resonance condition} is violated for the transitions of $V^{per}(t)$, %\fh{Comment on use of Localization here and comment of the use of \eqref{resonance condition}.}
\begin{equation}\label{form of A}
A_1(t) =  \sum_{i,w_1,w_2}A_{i,i+1}^{(w_1,w_2)}(t) \sigma_{i}^{(w_1)}\sigma_{i+1}^{(w_2)}
\end{equation}
with the bounds
\begin{equation}\label{bound on A}
\sup_{t,i,w_1,w_2} |A_{i,i+1}^{(w_1,w_2)}(t)| \le \delta^{-1}g/\nu
\end{equation}
(see Supplementary Material~II).
We compute 
\begin{align*}
&H_{3/2}(t) := P_1^{per,\dagger}(t) \left( H(t) - \id \frac{\dd}{\dd t} \right) P_1^{per}(t) \nonumber\\
%&= P_1^{per,\dagger}(t) \left( H_1^{(0)} + V_1^{per}(t) - \id \frac{\dd}{\dd t} \right) P_1^{per}(t) + P_1^{per,\dagger}(t) V_1^{res}(t) P_1^{per}(t) \nonumber\\
&=H^{(0)} + V^{res}(t) + \sum_{n\ge 1} \frac{\ad_{A_1(t)}^n}{n!} \left( \frac{n V^{per}(t)}{n+1}  + V^{res}(t) \right) \label{H 3/2}
\end{align*}
with $\ad_A(B)=[A,B]$ (to simplify the expression we pretended that $\ed^{A_1 (0)}=1$ \cite{footnote A1 non-zero}).
This expansion converges if $\delta^{-1} g/\nu$ is sufficiently small, see (2.7)-(2.10) in \cite{Imbrie}. 

From now on we assume that $\delta^{-1} g/\nu \ll 1$, so that, except for $V_1^{res} (t)$, all the time-dependent terms in $H_{3/2} (t)$ have been reduced by one factor $\delta^{-1} g/\nu$ at least.
We now define $P_1^{res}(t)$ as to get rid of $V_1^{res} (t)$. 
Thanks to \eqref{definition of P(t)}, there exists $P_1^{res}(t)$ so that
$$ P_1^{res,\dagger}(t)  \left( H^{(0)} + V^{res}(t) - \id \frac{\dd}{\dd t} \right) P_1^{res}(t)$$
is time-independent.
The main point is that, thanks to \eqref{probability resonance}, the Hamiltonian $H^{(0)} + V^{res}(t)$ acts non-trivially only on small and far between connected sets of bonds (the resonant spots).
Therefore, though $P_1^{res}(t)$ is not explicit, we know that it cannot ruin the localization, as it acts only inside the resonant spots. 
 
The first step is now completed, defining $H_2(t)$ via \eqref{H n+1 from H n}. 
%$$ H_2 (t)  =  P_1^{res,\dagger}(t) P_1^{per,\dagger}(t)   \left( H_1(t) - \id \frac{\dd}{\dd t} \right) P_1^{per}(t) P_1^{res}(t).$$
In order to iterate the scheme, we decompose again $H_2 (t) = H_2^{(0)} + V^{(2)}(t) $ with $H_2^{(0)} = \frac{1}{T}\int_0^T \dd t \, H_2(t)$, and we need to make sure that $H_2^{(0)}$ is MBL. 
As the diagonal part of $H_2^{(0)}$ is still given in leading approximation by $H^{(0)}$, with disorder strength $W$, and as the off-diagonal elements are at most of order $\delta^{-1} g^2 /\nu$ from the expression of $H_{3/2}(t)$, 
we find the condition $\delta^{-1}g^2 /\nu \ll W$. 
Dropping the artificial factor $\delta^{-1}$, the conditions for localization read
\begin{equation}\label{conditions on T and g}
g/\nu \ll 1, \quad   g^2 / \nu W \ll 1. 
\end{equation}
Unless the driving varies smoothly with time, we expect in general delocalization if one of these bounds get violated (see Supplementary Material III). 
%\fhB{Unless this results in a huge gain of place, I would prefer to keep a distinction between external citations and our own Supp. Mat.}

All the key points to proceed with the reduction of the time-dependent part of $H(t)$ at further and further scales did show up already at the first scale, 
and most of the technical work involved in the process can at this point be borrowed from \cite{Imbrie}; 
this includes the expansion of the perturbation in graphs, combinatorics estimates involved in the expansions of perturbative change of variables, the precise statement of resonance conditions, etc.\@
(see Supplementary Material IV for higher order resonances).
%\fhB{I moved here the sentence that was previously in the Supp. Mat. As we discovered with W.\@ while trying to read Imbrie in details, his text is surely not error-free, and a mathematician would not call this a proof 
%(though I still consider this as very serious work). For this reason, one should probably avoid too strong claims like ``convergence of the scheme is shown in the same way as Imbrie."}

%Though the implementation of the rest of the scheme is technically much more involved (see \cite{Imbrie}), this first step already contained the ideas needed to ``transpose" our understanding of time-independent MBL systems to driven Hamiltonians
%(see Supplementary Material IV). 

% SECTION
{\bf Localization at lower frequency.} 
We now consider the case where the driving contains only a few harmonics (e.g. $V_i (t) = \cos (2\pi \nu t) \overline V_i$ with $\overline V_i$ time-independent) 
and {we argue that delocalization occurs at frequencies which are lower than the threshold given by \eqref{conditions on T and g}.}
%that lower frequencies (i.e.\@ such that $\nu/g \ge 1$) are allowed before that the system delocalizes.  
 
%Next, we will argue that at sufficiently low driving frequencies, the properties of the Floquet operator $U(T)$ change dramatically compared to the high-frequency limit considered above: 
%$U(T)$ strongly mixes eigenstates of $H^{(0)}$, and the Floquet eigenstates become delocalized. 

\begin{figure}[htb]
\begin{center}
\includegraphics[width=0.9\columnwidth]{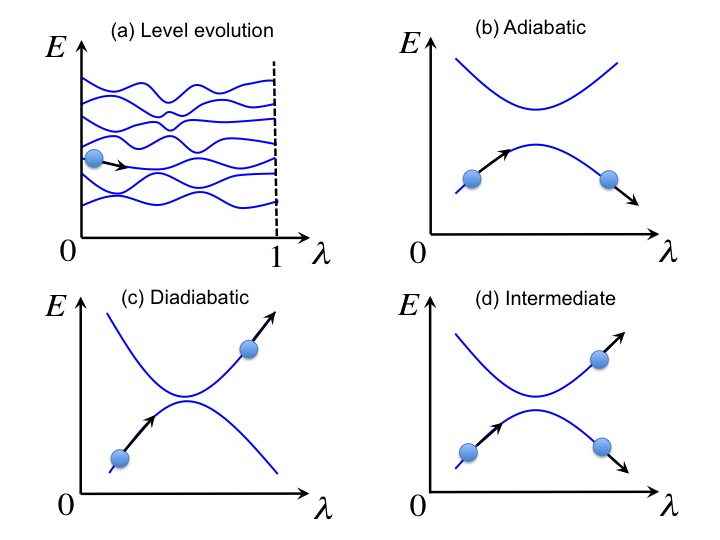}
\caption{ \label{Fig1} 
(a) Spectrum of the many-body localized system as a function of parameter $\lambda$. (b-d) Three kinds of level crossings: (b) Adiabatic, when the system follows instantaneous eigenstate, (c) Diabatic, when the system ends up in the original eigenstate, and (c) Intermediate, when the system is in a superposition of two states at long times.}
\end{center}
\end{figure}

Our argument relies on the analogy with the multi-level Landau-Zener problem. 
It is convenient to slightly modify the notations:
We introduce the dimensionless parameter $\lambda=\nu t$ and function $\hat V (\lambda) = V(\lambda)/g$, so that 
\be\label{eq:Hl}
H(\lambda)=H^{(0)}+g  \hat V(\lambda).
\ee
We assume that $g/W\ll 1$ is small enough so that each Hamiltonian $H(\lambda)$ is MBL for any $\lambda\in [0,1]$ with a localization length comparable to $\xi$ (the localization length of $H^{(0)}$).
% and so that the localization length of $H$. \fh{$\hat V$ is dimensionless.}
%\fh{We use $\xi$ for all localization lengths. We said twice $\xi \ll 1$. $g$ is small enough for all the $\xi$ to be of the same order.}
For the sake of exposition, let us split $\hat V=\hat V_{d}+\hat V_{od}$ such that $\hat V_{d}$ is the part of the perturbation that commutes with $H^{(0)}$. 
Then we write $\alpha$ and $ E_\alpha=E_\alpha(\lambda)$ for the eigenvectors and energies of $H^{(0)}+g\hat V_d(\lambda)$. As $\lambda$ goes through a cycle, these levels can cross, 
whereas the levels of $H(\lambda)$ have an avoided crossing (see Fig.~\ref{Fig1}(a)). 
The character of a pairwise level crossing is determined by (i) the matrix element of the operator $\hat V(\lambda)$ between the energy levels $|\alpha\ra, |\beta\ra$ that undergo the crossing, 
$M_{\alpha\beta}=\la \beta| \hat V_d(\lambda_c)|\alpha\ra$,  where $\lambda_c$ is the value of parameter $\lambda$ at which crossing takes place, 
and by (ii) the speed at which the crossing is passed: $v_{\alpha\beta}=\frac{d(E_\alpha(\lambda)-E_\beta (\lambda))}{d\lambda}  \nu$.
In the Landau-Zener problem (crossing of just two levels), the transition amplitude is given by (see e.g.\@ \cite{Shytov04}): 
\be\label{eq:S11}
S_{\alpha\to \alpha}=\exp(-C_{\alpha\beta}), \,\, C_{\alpha\beta}\equiv \pi \frac{|M_{\alpha\beta}|^2}{v_{\alpha\beta}}, 
\ee
and therefore one can distinguish three regimes: 
(i) {\it adiabatic}, when parameter $C_{\alpha\beta}\gg 1$; in this case, the system ends up in eigenstate $\beta$ after the crossing is passed, and the probability to stay in the ``excited" state $\alpha$ is exponentially small; 
(ii) {\it diabatic}, when $C_{\alpha\beta}\ll 1$; in this case, the system stays in state $\alpha$;
(iii) {\it intermediate}, when $C_{\alpha\beta}\sim 1$; in this case, the system ends up in a superposition of states $\alpha$ and $\beta$ at long times, with approximately similar weights. The three regimes are illustrated in Fig.~\ref{Fig1}(b-d). 

%% REMARK: \bar{x} became x_1, and x^* became x_2

As we will now argue, in our problem the relevant crossings, which lead to delocalization at low frequency, occur between levels that differ only by a small number of LIOMs. 
Let us consider two levels $\alpha$, $\beta$, which have different values of local integrals of motion only in a region $R$ of size $x\ll L$. 
We first show that there is a scale $x_1$, such that at $x\ll x_1$ the crossings between states which differ in region $R$ are very rare, while at $x\gg x_1$ there are many such crossings. 
There are $2^x$ different levels which have different value of LIOMs in the region $R$ and are identical outside $R$, 
and an overwhelming majority of these levels lives in a band of width of order $W \sqrt x$ \cite{footnotenormal}.
Therefore, the typical level spacing for this group of levels can be estimated as: 
\be\label{eq:level_spacing}
\Delta(x)\approx  W\frac{\sqrt{x}}{2^x}.
\ee
On the other hand, the typical change of {energy difference} between two levels $\alpha, \beta$, whose LIOMs differ only in the region $R$, can be estimated as
\be\label{eq:en_diff}
\delta E_{\alpha\beta} (x)\sim  g  | \la \alpha|  \hat V |\alpha\ra -\la \beta| \hat V |\beta \ra |  \sim g \sqrt{x},
\ee
when $\delta\lambda \sim 1$, and where we used $\norm \hat V_i \norm \sim 1$. 
If $\delta E_{\alpha\beta}(x)$ is much smaller than the level spacing $\Delta(x)$, the levels in this group typically do not cross. 
In the opposite limit, $\delta E_{\alpha\beta}\gg \Delta(x)$, there are multiple level crossings of this kind. 
The scale $x_1$ can therefore be estimated from the condition $\delta E_{\alpha\beta}(\bar{x})\approx \Delta(x_1)$, which gives:
\be\label{eq:xbar}
x_1=\log _2 \frac{W}{g}. 
\ee
At $x\gtrsim x_1$, each level $\alpha$ therefore crosses multiple other levels which differ from $\alpha$ by changing values of some or all LIOMs in (any) region of size $x$. 

Next, let us understand the character of crossings between levels $\alpha, \beta$ that have different LIOMs only in a region $R$ or size $x$ (assuming that such a crossing is encountered as $\lambda$ is varied). 
First, we estimate the speed at which the crossing is passed: $v_{\alpha\beta}(x)\sim \delta E_{\alpha\beta} \nu \sim g \nu  \sqrt{x}$. 
Second, we note that the typical matrix element of a local operator between two MBL eigenstates which differ in region $R$, is given by: 
\be\label{eq:matrix_element}
M_{\alpha\beta}(x) \sim g  \la \alpha| \hat V|\beta \ra\sim g \sqrt{x} e^{-x/\xi}.
\ee

The value of the parameter $C_{\alpha\beta}(x)$ characterizing the crossing is then given by: 
\be\label{eq:Cabx}
C_{\alpha\beta}(x)\sim \frac{g\sqrt{x}}{\nu}  e^{-2x/\xi}. 
\ee
The crossing is in the intermediate regime (the two crossing levels mix strongly at long times) at scale $x_2$, which can be estimated from the relation $C_{\alpha\beta}(x_2)\sim 1$: 
\be\label{eq:x*}
x_2 \approx \frac{\xi}{2} \log \frac{g}{\nu}.
\ee
At $x\gg x_2$, (nearly) all crossings are in the diabatic regime, while at $x\ll x_2$ crossings are adiabatic. 

The properties of the Floquet operator, most importantly the way it mixes states with very different spatial structure, depend on the relation between length scales $x_1,x_2$ given by (\ref{eq:xbar},\ref{eq:x*}). 
If $x_2\gg x_1$, during one period of driving, each level experiences multiple crossings which are in the intermediate {or adiabatic} regime. 
This means that the operator $U(T)$ cannot be considered a small perturbation of $U_{g=0}(T)=\exp(-\iu T H^{(0)})$, as it changes the values of most LIOMs. 
Hence, in this case we expect that the eigenstates of $U(T)$ are  delocalized.

\begin{figure}[htb]
\begin{center}

\begin{tikzpicture}[scale=0.8]

\draw[fill=green!20] (-2,-2) -- (2,-2) -- (2,2) -- (-2,2) -- (-2,-2);

\draw[fill=pink!40] (0,0) circle (1.5);

\draw[dotted] (0,0) circle (1);

\draw[<->,>=latex] (0,0)--(-1.06,1.06);
\draw[<->,>=latex] (0,0)--(0.707,0.707);

\draw (0,-0.55) node[below] {\tiny No crossing};
\draw (0,2) node[below] {\tiny Diabatic crossings};
\draw (0.6,0.5) node[below] {\tiny $x_2$};
\draw (-0.5,0.5) node[below] {\tiny $x_1$};

\begin{scope}[xshift=5cm]
\draw[fill=green!20] (-2,-2) -- (2,-2) -- (2,2) -- (-2,2) -- (-2,-2);
\draw[fill=white] (0,0) circle (1.5);
\draw[fill=pink!40] (0,0) circle (1);

\draw[<->,>=latex] (0,0)--(1.06,1.06);
\draw[<->,>=latex] (0,0)--(-0.707,0.707);

\draw (0,0) node[below] {\tiny No crossing};
\draw (0,-0.9) node[below] {\tiny Adiabatic};
\draw (0,2) node[below] {\tiny Diabatic crossings};
\draw (0.6,0.5) node[below] {\tiny $x_2$};
\draw (-0.5,0.5) node[below] {\tiny $x_1$};

\end{scope}

%\draw[->,>=latex] (-0.3,0) -- (6,0);
%\draw[->,>=latex] (0,-0.3) -- (0,4);

%\draw[dotted] (1,0) -- (1,1) -- (0,1);

%\draw (1,0) -- (1,-0.1);
%\draw (0,1) -- (-0.1,1);

%\draw (1,0) node[below]{1};
%\draw (0,1) node[left]{1};

%\draw[ultra thick,Red] plot [domain=0.265:5.8,smooth] (\x, {1/\x}); 
%\draw[ultra thick,Red] plot [domain=0.13:1,smooth] (\x, {1/\x^(2/3)}); 

%\draw  (0,3.5) node[left] {{$W/g$}};
%\draw  (5.5,0) node[below] {{$\nu/g$}};

%\draw  (0.9,0.15) node[above] {Thermal};
%\draw  (3,2) node[above] {Localized};

\end{tikzpicture}

\caption{ \label{Figure: two scales} 
Scales $x_1$ and $x_2$ (here $d=2$ for visualization). 
Left panel: at high frequency, when \eqref{eq:nu_critical} holds, all the crossings are typically diabatic. 
Right panel: at lower frequency, when \eqref{eq:nu_critical} is violated, adiabatic crossings typically appear.}
\end{center}
\end{figure}
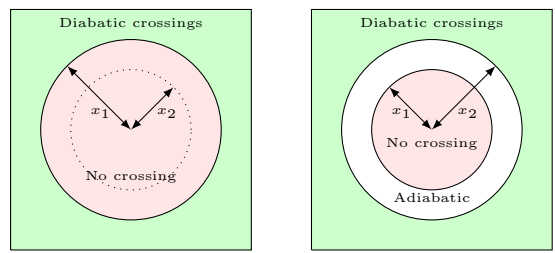

From (\ref{eq:xbar},\ref{eq:x*}), we deduce that the condition $x_2 \ll x_1$ for localization is written in terms of the frequency $\nu$ as 
% Condition for DELOcalization:  $x_*\gg \bar{x}$ i.e. $x_2 \gg x_1$
\be\label{eq:nu_critical}
\frac{g}{\nu} \Big(\frac{g}{W}\Big)^{1/a} \ll 1, \quad \text{with}\quad a = \frac{\xi \log 2}{2}.
%\nu\ll \nu_*, \,\, \nu_*=g \left(  \frac{g}{W}\right)^{1/a}, \,\,\,  a\equiv \frac{\xi \log 2}{2},  
\ee
%Since we assume $g/W \ll 1$, it holds that $a \sim \xi \sim (\ln (W/g))^{-1} < 1$. 
By the above reasoning, we expect delocalization once condition \eqref{eq:nu_critical} gets violated.
See Fig.~\ref{Figure: two scales}. 
%In its present state, our reasoning does not yet allow to firmly conclude that \eqref{eq:nu_critical} guaranties localization. 
%Instead, once \eqref{eq:nu_critical} is violated, we expect the system to delocalize. \fh{Idea ok, but formulation not optimal.}
%\fh{If $h<< g$ still with }
%Thus, at sufficiently low driving frequency, the system {\it always delocalizes}.

%As we showed above, in the limit of high frequency (Eqs.(\ref{eq: 1st condition},\ref{eq: 2d condition})), the system is in the MBL phase. 
%Therefore, we expect a localization-delocalization transition to take place at some critical frequency $\nu_c$.   We conjecture that at $g\ll W$ (the assumption under which the analogy with Landau-Zener problem can be invoked), the  transition takes place when $x_*\approx \bar{x}$, that is, $\nu_c\approx \nu_*$, although we were not able to prove this fact. Further, in the limit of large driving strength, $g\gg W$, the perturbation theory shows that MBL persists at $\nu \gg \sqrt{gW}$. The perturbation theory breaks down at $\nu \sim \sqrt{gW}$, and we conjecture that in this limit delocalization occurs at $\nu_c\sim \sqrt{gW}$. Combining these two results, we propose a phase diagram in Fig.\ref{Fig3}. 

{\bf Discussion.} 
In summary, we have provided strong analytical evidence that many-body localization persists under periodic driving, if the driving frequency is high enough. 
The MBL phase in driven systems is characterized by the existence of an MBL (time-independent) effective Hamiltonian, implying thus 
(i) the existence of an extensive number of local conservation laws~\cite{Serbyn13-2,Huse13,Ros14,Chandran14}; 
(ii) area-law for all, but an exponentially small fraction of Floquet eigenstates~\cite{Serbyn13-2,bauer,marienverstraete,footnotearea}; 
(iii) logarithmic spreading of entanglement entropy of initial product states~\cite{Prosen08,Moore12,Vosk13,Serbyn13-1,Serbyn13-2,Huse13}. 
At sufficiently low driving frequencies, the system undergoes a transition into the delocalized phase.
%\wdr{Do we mean anything concrete by this 'almost certainly'?  Does it refer to the possibility of delocalized but non-ergodic phases? If so, I would definitely drop it because I think there is no sense in which our analysis would predict more strongly deloc than ergodic, or opposite}
%In the future, it will be interesting to explore the nature of the MBL-delocalization transition in driven systems, and in particular, to understand whether it belongs to the same universality class as the MBL-delocalization transition for static Hamiltonians. 

We note that our results are in agreement with two previous recent studies~\cite{Lazarides14-2,Ponte14-2}, which provided qualitative arguments and numerical evidence for the existence of the MBL phase at large driving frequency and delocalization at small frequency.

{\bf Acknowledgements.} We thank A.~Polkovnikov, P.~Ponte, Z.~Papi\'c, Y.~Wan and an anonymous referee for helpful discussions and suggestions. 
D.A.\@ acknowledges support by Alfred Sloan Foundation.  
W.D.R.\@ thanks the DFG (German Research Fund) and the Belgian Interuniversity Attraction Pole (P07/18 Dygest) for financial support. 
Both F.H.\@ and W.D.R.\@ acknowledge the support of the ANR grant JCJC.

%
%
%
%
%%%% SUPPLEMENTARY MATERIAL
%
%
%
%
\pagebreak
\setcounter{figure}{0}
\setcounter{page}{1}
\renewcommand{\thefigure}{S\arabic{figure}}

\begin{widetext}
\begin{center}
\section{Supplementary Material: A theory of many-body localization in periodically driven systems}

Content of the Supplementary Material:
Sections I and II contain straightforward computations that allow us to derive the relations \eqref{probability resonance}, \eqref{form of A} and \eqref{bound on A} in the main text. 
Section III bridges a link between our two conditions \eqref{conditions on T and g} and \eqref{eq:nu_critical} for localization. In Section IV, we show how to estimate the probability of resonances in higher orders.

\end{center}
\end{widetext}

% Appendix I
\textbf{I. Derivation of \eqref{probability resonance}.}
A Gaussian approximation yields  
\begin{align*}
&\Proba ((i,i+1) \in \Res) \\
&\le \sum_{k\ne 0}\sum_{a,b\in\{0,\pm 2Ê\}}  \Proba (|a \epsilon_i + b\epsilon_{i+1} - 2\pi k \nu| \le \delta \nu) \\
&\sim \sum_{k\ne 0;a,b} \int_{-\infty}^{+\infty}  \frac{\dd z}{W}\, \ed^{-\frac{(z-2 \pi k\nu)^2}{(\delta \nu)^2}} \ed^{-\frac{(z-(a+b)E_0)^2}{W^2}}\\
&\sim \sum_{k\ne 0;a,b} \delta \nu \frac{\ed^{-\frac{(2 \pi k\nu - (a+b)E_0)^2}{(\delta \nu)^2 + W^2}}}{\sqrt{(\delta \nu)^2 + W^2}}\\
& = \delta \sum_{a,b} \sum_{k\ne 0}  \frac{\ed^{-\frac{(2\pi k - (a+b)E_0/\nu)^2}{\delta^2 + (W/\nu)^2}}}{\sqrt{\delta^2 + (W/\nu)^2}} \le \Const \delta, 
\end{align*}
where the last bound follows from the fact that $k\ne 0$ and $E_0 \sim W$, so that the sum over $k$ can be approximated by an integral of a probability density. 

Though we derived \eqref{probability resonance} through a Gaussian approximation for simplicity, it is much more general; 
the important point is that the distribution of $\epsilon_i$ does not force the energies to be such that $a\epsilon_i + b\epsilon_j$ is typically near a value of the form $2\pi k \nu$ for some $k\ne 0$, for which the key ingredient was here $E_0\sim W$. 
%Let us fix $a,b\in \{Ê0,\pm 1\}$. 
%If $|(a+b)E_0/\nu| >1/2$, the sum over $k$ can be bounded by the integral of a probability density.  
%If $|(a+b)E_0/\nu| \le1/2$, we first use the constraint $k\ne 0$ to get the bound $(2 \pi k - (a+b)E_0/\nu)^2 \ge k^2 /2$, and then bound the sum by the integral of a probability density. 
%In both cases, summing over $a,b \in \{Ê0,\pm 1 \}$, we obtain \eqref{probability resonance}.
Finally, it may appear as a surprise that $W$ is not involved in the bound \eqref{probability resonance}. 
This is a consequence of our assumption $\frac{1}{T}\int_0^T \dd t \, V(t) = 0$, implying that we can discard $k=0$ in \eqref{probability resonance}.
The condition on $W$ appears later, when requiring that $H_2^{(0)}$ is MBL in \eqref{conditions on T and g}.

% Appendix II
\textbf{II. Derivation of \eqref{form of A}-\eqref{bound on A}.}
The proof of these relations relies crucially on the fact that condition \eqref{resonance condition} is violated, and that $H_0$ is MBL, i.e.\@ that a local operator couples only a few close LIOMs
(due to our simplifications, this is here obvious since both the eigenstates of $H^{(0)}$ and the perturbation terms $V_i(t)$ are strictly local).
In terms of the matrix elements, \eqref{perturbative equation} reads
\begin{align*}
&\Big( \Delta H^{(0)}_{\eta,\eta'} - \id \frac{\dd }{ \dd t} \Big) \langle \eta' |ÊA_1(t) | \eta \rangle = \langle \eta' |ÊV^{per}(t) |Ê\eta \rangle, \\
& \langle \eta' |ÊA_1(t) |Ê\eta\rangle =   \langle \eta' |ÊA_1(t+T) |Ê\eta\rangle. 
\end{align*}
When $ \Delta H^{(0)}_{\eta,\eta'} = 0$, i.e.\@ for $\eta = \eta'$, the solution exists thanks to the condition $\frac{1}{T}\int_0^T \dd t \, V^{per}(t)=0$ and is given by 
$$\langle \eta |ÊA_1(t) |Ê\eta\rangle = \id \int_0^t \dd s \,  \langle \eta |ÊV^{per}(s) |Ê\eta \rangle $$
where we have chosen the initial condition $\langle \eta' |ÊA_1(0) |Ê\eta\rangle =0$. 
When $ \Delta H^{(0)}_{\eta,\eta'} \ne 0$, i.e.\@ when $\eta \ne \eta'$, the solution is given by 
\begin{align*}
&\langle \eta' |ÊA_1(t) |Ê\eta\rangle = \id \int_0^t \dd s \, \ed^{- \id \Delta H^{(0)}_{\eta,\eta'} ((t-s))}   \langle \eta' |ÊV^{per}(s) |Ê\eta \rangle \\
& + \id \frac{\ed^{-\id \Delta H^{(0)}_{\eta,\eta'} t}}{ 1 - \ed^{-\id \Delta H^{(0)}_{\eta,\eta'} T}}  \int_0^T \dd s \, \ed^{-\id  \Delta H^{(0)}_{\eta,\eta'} (T-s) }   \langle \eta' |ÊV^{per}(s) |Ê\eta \rangle .
\end{align*}
Using again the condition $\frac{1}{T}\int_0^T \dd t \, V^{per}(t)=0$, we may replace $\ed^{-\id  \Delta H^{(0)}_{\eta,\eta'} (T-s) }$ by $\ed^{-\id  \Delta H^{(0)}_{\eta,\eta'} (T-s) }  - 1$ in this last integral, and use the bound 
$$ \left| \frac{\ed^{-\id  \Delta H^{(0)}_{\eta,\eta'} (T-s) }  - 1}{1 - \ed^{-\id \Delta H^{(0)}_{\eta,\eta'} T}} \right| \le \frac{2}{\min_{k\in\Z_0}\{ | T\Delta H^{(0)}_{\eta,\eta'} - 2\pi k | \} }.$$
Since the resonance condition \eqref{resonance condition} is always violated by the matrix elements of $V^{per}(t)$, we obtain the bound 
$$\sup_t | \langle \eta' |ÊA_1(t) |Ê\eta \rangle |  \le \Const \delta^{-1}T \sup_t | \langle \eta' |ÊV^{per}(t) |Ê\eta \rangle | \quad \forall \eta, \eta'.  $$
From there, since \eqref{perturbative equation} is linear, we recover  \eqref{form of A}-\eqref{bound on A}.

% Appendix III
\textbf{III. Smooth vs non-smooth $V(t)$.}
As it is seen \emph{a posteriori} from \eqref{eq:nu_critical}, the conditions \eqref{conditions on T and g} are not always optimal conditions for localization. 
This comes from the fact that, when the driving $V(t)$ varies smoothly with time, i.e.\@ is concentrated on a few harmonics, the way we control the solutions of \eqref{perturbative equation} can be improved in some regimes. 
In view of our estimates based on the Landau-Zener crossings, we would like to consider smaller values of $\nu$, violating the contraint $g/\nu < 1$, while requiring now $g/W \ll 1$. 
Here we show how our renormalization scheme could be adapted to deal with this case, providing a direct argument for \emph{localization} in this regime, 
while the more physical arguments leading to the condition \eqref{eq:nu_critical} ensured primarily that the system \emph{delocalizes} once this latter condition is violated. 
We keep however the discussion at the level of general ideas (in particular we do not show how to recover \eqref{eq:nu_critical}), postponing a more throughout investigation for further works.

Let us move to the Fourier variables in \eqref{perturbative equation}: 
$$ G(k) = \frac{1}{T}\int_0^T \dd t \, \ed^{2 \id \pi k \nu t} G (t), \quad k\in \Z,$$ 
for $G=A_1$ and $G=V$ (we work directly with $V$ instead of $V^{per}$ as it would not make sense to maintain this distinction at this level). 
Smoothness (more precisely analyticity) for $V(t)$ in the time domain reads $ |V(k)| \sim g\ed^{-c|k|} $ for some $c> 0$ in the frequency domain. 
Eq.~\eqref{perturbative equation} becomes 
$$[H^{(0)},A_1(k)] - 2\pi k\nu A_1 (k)  =V (k), \quad k\ne 0 $$
(remember that $V(k=0) = \frac{1}{T}\int_0^T \dd t \, V(t) = 0$).
Thus, for any $\eta,\eta'$, 
\begin{equation}\label{solution in Fourier}
\langle \eta' |ÊA_1 (k) |Ê\eta \rangle = \frac{\langle \eta' |ÊV(k) |Ê\eta\rangle }{\Delta H^{(0)}_{\eta,\eta'} -2\pi k \nu}. 
\end{equation}

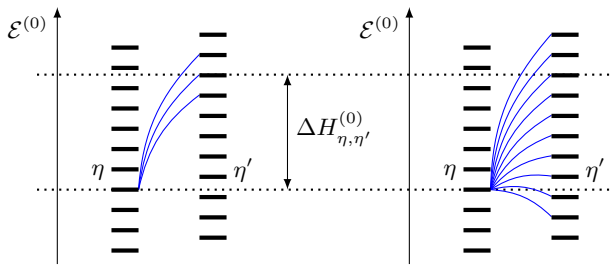
\begin{figure}[htb]
\begin{center}

\begin{tikzpicture}[scale=0.9]

%%%%%%%% Smooth
%\draw (-0.1,0) -- (3.2,0); \draw (-0.1,0) node[left]{$0$};
\draw[->,>=latex] (0,-1.5) -- (0,2.3);
\draw (0.6,0.1) node[below]{$\eta$};
\draw (2.75,0.2) node[below]{$\eta'$};
\draw (0,2) node[left] {$\mathcal E^{(0)}$};

\draw[<->,>=latex] (3.4,-0.4) -- (3.4,1.3); \draw (3.4,0.5) node[right]{$\Delta H^{(0)}_{\eta,\eta'} $};
\draw[dotted,thick] (-0.3,-0.4) -- (8.2,-0.4);
\draw[dotted,thick] (-0.3,1.3) -- (8.2,1.3);

\draw[color=blue] (1.2,-0.4) to [bend left = 20] (2.1,1);
\draw[color=blue] (1.2,-0.4) to [bend left = 20] (2.1,1.3);
\draw[color=blue] (1.2,-0.4) to [bend left = 20] (2.1,1.6);

\draw[ultra thick] (0.8,-1.3) -- (1.2,-1.3);
\draw[ultra thick] (0.8,-1) -- (1.2,-1);
\draw[ultra thick] (0.8,-0.7) -- (1.2,-0.7);
\draw[ultra thick] (0.8,-0.4) -- (1.2,-0.4);
\draw[ultra thick] (0.8,-0.1) -- (1.2,-0.1);
\draw[ultra thick] (0.8,0.2) -- (1.2,0.2);
\draw[ultra thick] (0.8,0.5) -- (1.2,0.5);
\draw[ultra thick] (0.8,0.8) -- (1.2,0.8);
\draw[ultra thick] (0.8,1.1) -- (1.2,1.1);
\draw[ultra thick] (0.8,1.4) -- (1.2,1.4);
\draw[ultra thick] (0.8,1.7) -- (1.2,1.7);

\begin{scope}[xshift=1.3cm,yshift=0.19cm]
\draw[ultra thick] (0.8,-1.3) -- (1.2,-1.3);
\draw[ultra thick] (0.8,-1) -- (1.2,-1);
\draw[ultra thick] (0.8,-0.7) -- (1.2,-0.7);
\draw[ultra thick] (0.8,-0.4) -- (1.2,-0.4);
\draw[ultra thick] (0.8,-0.1) -- (1.2,-0.1);
\draw[ultra thick] (0.8,0.2) -- (1.2,0.2);
\draw[ultra thick] (0.8,0.5) -- (1.2,0.5);
\draw[ultra thick] (0.8,0.8) -- (1.2,0.8);
\draw[ultra thick] (0.8,1.1) -- (1.2,1.1);
\draw[ultra thick] (0.8,1.4) -- (1.2,1.4);
\draw[ultra thick] (0.8,1.7) -- (1.2,1.7); 
\end{scope}

%%%%%%%%% Non-smooth
\begin{scope}[xshift=5.2cm,yshift=0cm]
%\draw (-0.1,0) -- (3.2,0); \draw (-0.1,0) node[left]{$0$};
\draw[->,>=latex] (0,-1.5) -- (0,2.3);
\draw (0.6,0.1) node[below]{$\eta$};
\draw (2.75,0.2) node[below]{$\eta'$};
\draw (0,2) node[left] {$\mathcal E^{(0)}$};

\draw[color=blue] (1.2,-0.4) to [bend left = 20] (2.1,-0.8);
\draw[color=blue] (1.2,-0.4) to [bend left = 20] (2.1,-0.5);
\draw[color=blue] (1.2,-0.4) to [bend left = 20] (2.1,-0.2);
\draw[color=blue] (1.2,-0.4) to [bend left = 20] (2.1,0.1);
\draw[color=blue] (1.2,-0.4) to [bend left = 20] (2.1,0.4);
\draw[color=blue] (1.2,-0.4) to [bend left = 20] (2.1,0.7);
\draw[color=blue] (1.2,-0.4) to [bend left = 20] (2.1,1);
\draw[color=blue] (1.2,-0.4) to [bend left = 20] (2.1,1.3);
\draw[color=blue] (1.2,-0.4) to [bend left = 20] (2.1,1.6);
\draw[color=blue] (1.2,-0.4) to [bend left = 20] (2.1,1.9);

\draw[ultra thick] (0.8,-1.3) -- (1.2,-1.3);
\draw[ultra thick] (0.8,-1) -- (1.2,-1);
\draw[ultra thick] (0.8,-0.7) -- (1.2,-0.7);
\draw[ultra thick] (0.8,-0.4) -- (1.2,-0.4);
\draw[ultra thick] (0.8,-0.1) -- (1.2,-0.1);
\draw[ultra thick] (0.8,0.2) -- (1.2,0.2);
\draw[ultra thick] (0.8,0.5) -- (1.2,0.5);
\draw[ultra thick] (0.8,0.8) -- (1.2,0.8);
\draw[ultra thick] (0.8,1.1) -- (1.2,1.1);
\draw[ultra thick] (0.8,1.4) -- (1.2,1.4);
\draw[ultra thick] (0.8,1.7) -- (1.2,1.7);

\begin{scope}[xshift=1.3cm,yshift=0.19cm]

\draw[ultra thick] (0.8,-1.3) -- (1.2,-1.3);
\draw[ultra thick] (0.8,-1) -- (1.2,-1);
\draw[ultra thick] (0.8,-0.7) -- (1.2,-0.7);
\draw[ultra thick] (0.8,-0.4) -- (1.2,-0.4);
\draw[ultra thick] (0.8,-0.1) -- (1.2,-0.1);
\draw[ultra thick] (0.8,0.2) -- (1.2,0.2);
\draw[ultra thick] (0.8,0.5) -- (1.2,0.5);
\draw[ultra thick] (0.8,0.8) -- (1.2,0.8);
\draw[ultra thick] (0.8,1.1) -- (1.2,1.1);
\draw[ultra thick] (0.8,1.4) -- (1.2,1.4);
\draw[ultra thick] (0.8,1.7) -- (1.2,1.7); 

\end{scope}

\end{scope}

%\draw[dotted] (1,0) -- (1,1) -- (0,1);

%\draw (1,0) -- (1,-0.1);
%\draw (0,1) -- (-0.1,1);

%\draw (1,0) node[below]{1};
%\draw (0,1) node[left]{1};

%\draw[ultra thick,Red] plot [domain=0.265:5.8,smooth] (\x, {1/\x}); 
%\draw[ultra thick,Red] plot [domain=0.13:1,smooth] (\x, {1/\x^(2/3)}); 

%\draw  (0,3.5) node[left] {{$W/g$}};
%\draw  (5.5,0) node[below] {{$\nu/g$}};

%\draw  (0.9,0.15) node[above] {Thermal};
%\draw  (3,2) node[above] {Localized};

\end{tikzpicture}

\caption{ \label{figure: smooth vs non-smooth} 
Horizontal lines represent the values of $\mathcal E^{(0)}(\eta,k) = H^{(0)}_\eta - 2 \pi k \nu$ for all $k\in \Z$, where $H^{(0)}_\eta = \langle \eta |ÊH^{(0)}|\eta\rangle$. 
Left panel: a smooth driving $V(t)$ connects only close harmonics, the shift being due to $\Delta H^{(0)}_{\eta,\eta'} \sim W$. 
Right panel: if $V(t)$ is not smooth, harmonics are connected even if they are far appart, leading to resonant couplings. 
}
\end{center}
\end{figure}

Let us first consider the case $\eta \ne \eta'$, so that typically $\Delta H^{(0)}_{\eta,\eta'} \sim W$. 
As long as $\nu \gtrsim W$, there is no space for improvement here. 
However, for $V$ smooth, the situation changes once $\nu \ll W$: 
$$\left|  \frac{\langle \eta' |ÊV(k) |Ê\eta\rangle }{\Delta H^{(0)}_{\eta,\eta'} -2\pi k \nu}  \right| \sim \frac{ g \, \ed^{-c |k|}}{| W - 2\pi k \nu |} \sim \frac{g}{W}$$
where we set $c\sim 1$ for simplicity.
Therefore, in this case, the (non-valid) condition $g/\nu \ll 1$ can be replaced by the (valid) condition $g/W \ll 1$, as we expected.
The difference between smooth and non-smooth drivings is illustrated on Figure \ref{figure: smooth vs non-smooth}. 

We are still left with the diagonal elements ($\eta = \eta'$). 
For them, $\Delta H^{(0)}_{\eta,\eta} = 0$, and the r.h.s.\@ of \eqref{solution in Fourier} behaves like $g/\nu$, which can be very large, independently of the value of $g/W$. 
We notice however that the resonances in the on-site terms do not need to entail delocalization. 
Indeed, for any diagonal Hamiltonian $D(t)$, eq.~\eqref{canonical transformation} can be solved with $D_* = \frac{1}{T}\int_0^T \dd s\,  D(s)$ and 
\begin{equation}\label{non perturbative Pt appendix}
P(t) = \ed^{-i \int_0^t \dd s \, (D(s) - D_*)}. 
\end{equation}
Taking for $D(t)$ the diagonal part of $V(t)$, we conclude that the corresponding matrix $P(t)$, though non-perturbative, is diagonal, local and preserves the product structure, hence the localization. 

If we were giving a precise description here, then there is a catch in the above argumentation, if not correctly interpreted.
Indeed, at low frequency, $P(t)$ defined by \eqref{non perturbative Pt appendix} does not inherit the smoothness of $D(t)$, 
i.e.\@ even if $D(k)$ is non-vanishing only for a few low harmonics $k$, $P(k)$ starts only decaying for $k\gtrsim g/\nu$.  
As the non-perturbative rotation matrix given by \eqref{non perturbative Pt appendix} affects also the non-diagonal matrix elements, it is not at all clear that our reasoning could be iterated starting from the next step of the scheme. 
Let us simply  point our here that this difficulty is only apparent and can be bypassed by performing the on-site non-perturbative rotations of the type \eqref{non perturbative Pt appendix} only at the right scale of the scheme, after several rotations of the type \eqref{solution in Fourier} have significantly driven down $V(t)$
(we postpone a more detailed explanation of this to a further work).

% Appendix IV
\textbf{IV. Higher order resonances.}
As this is truly at the heart of the localization phenomenon, we check here explicitly that resonances rarefy quickly as higher and higher order couplings are considered.  
A $n^{\mathrm{th}}$-order coupling $V^{(n)}_i (t)$ is of order $(g/\nu)^{n-1}g$ and connects at most $n$ consecutive spins.  
A resonance occurs if, for two eigenstates $|\alpha\rangle$ and $| \beta \rangle$ of the effective Hamiltonian at that scale 
(i.e.\@ the Hamiltonian $H_m^{(0)}$ for some $m=m(n)$)
%, where the scale $m$ depends on the order $n$ in a way fixed by the scheme, roughly $m \sim \log n$, see \cite{Imbrie}) 
that differ only in a box of size $n$, we have 
$$ \frac{(g/\nu)^{n-1} g}{|\Delta E_{\alpha,\beta} - 2\pi k \nu|} \gtrsim 1 \quad \text{for some }k\in \Z_0.$$  
{As resonances never occur for $|\alpha\rangle = |\beta\rangle$ in the regime $g/\nu \ll 1$, we assume $|\alpha\rangle \ne | \beta\rangle$, where we do not expect any help from the extra condition $k\ne 0$.
Now, the quantity $\min_{k\in \Z} |E_\beta - E_\alpha - 2\pi k \nu |$ can be visualized as the distance between $E_\alpha$ and $E_\beta$, projected on a circle of circumference $2\pi \nu$ (i.e.\@ their difference modulo $2\pi \nu$). 
We need thus to consider $2^n$ eigenvalues on this circle, so that the average level spacing is of order $2^{-n}\nu$} (for a time-dependent problem, it is of order $ 2^{-n} \sqrt n W$, which is not better for any practical purpose)
and the smallest level spacing can be estimated by $\ed^{-cn} {\nu}$ for some $c > 0$ with high probability 
(to be precise: we make the assumption that there is no conspiracy in the system that produces  anomalously small spacings, cfr.\ the analogous LLA assumption in \cite{Imbrie}).
Therefore, if $g/\nu$ is small enough, we expect resonances to become quickly very atypical at higher scales, hence localization.


\begin{thebibliography}{99}
\bibitem{footnotearea}
Let us consider LIOMs that are obtained from the physical spins by a quasilocal unitary transformation that is generated by a time-dependent Hamiltonian (the operator $\id A$ in our case). 
Then the work in \cite{marienverstraete} shows that such quasilocal unitaries preserve the area-law property.
\bibitem{remarkenergyloc}  
Remark that, for energy localization, it is already sufficient to find $H_*$ that can be written as a sum of local terms, even if $H_*$ itself is ergodic, i.e.\@ not MBL.
However, the only instance where we know that this happens ($H_*$ local but not MBL) is for free fermions or bosons at $\nu$ larger than the bandwidth.
\bibitem{neglect-original-res} 
This amounts {essentially} to neglecting local resonant spots in $H^{(0)}$. This is a posteriori justified since these spots should be treated in the same way as resonances induced by the driving, which are discussed in the present paper.
\bibitem{footnotenormal}
We assume a Gaussian density of states, though the power of $x$ here does not influence our conclusion. For a mathematical result on Gaussian density of levels, see e.g.\ \cite{keatingnormal}
\end{thebibliography}
\end{document}